\documentclass[lettersize,journal]{IEEEtran}

\usepackage{amsmath,amsfonts}
\usepackage{algorithmic}
\usepackage{array}
\usepackage[caption=false,font=normalsize,labelfont=sf,textfont=sf]{subfig}
\usepackage{textcomp}
\usepackage{stfloats}
\usepackage{url}
\usepackage{verbatim}
\usepackage{graphicx}
\usepackage{cite}
\usepackage{xspace}
\usepackage[svgnames]{xcolor}
\usepackage{multirow}
\usepackage{booktabs}
\usepackage{cleveref}
\usepackage{tcolorbox}
\usepackage{enumitem}
\usepackage[linesnumbered,ruled,vlined]{algorithm2e} 

\usepackage{listings}
\lstdefinelanguage{diff}{
  morecomment=[f][\color{dimgray}]{@@},
  morecomment=[f][\color{cadmiumgreen}]{+\ },
  morecomment=[f][\color{brickred}]{-\ },
}

\def\name{\textsc{Fonte}\xspace}
\def\product{SAP HANA\xspace}

\definecolor{dimgray}{rgb}{0.41, 0.41, 0.41}
\definecolor{brickred}{rgb}{0.8, 0.25, 0.33}
\definecolor{cadmiumgreen}{rgb}{0.0, 0.42, 0.24}
\definecolor{dkgreen}{rgb}{0,0.6,0}
\definecolor{mauve}{rgb}{0.58,0,0.82}
\definecolor{gray}{rgb}{0.4,0.4,0.4}
\definecolor{darkblue}{rgb}{0.0,0.0,0.6}
\definecolor{lightblue}{rgb}{0.0,0.0,0.9}
\definecolor{cyan}{rgb}{0.0,0.6,0.6}
\definecolor{darkred}{rgb}{0.6,0.0,0.0}

\begin{document}

\title{Identifying Bug Inducing Commits by Combining Fault Localisation and Code Change Histories}

\author{Gabin~An,
Jinsu~Choi,
Jingun~Hong,
Naryeong~Kim,
Shin~Yoo
\IEEEcompsocitemizethanks{
\IEEEcompsocthanksitem Gabin An, Jinsu Choi, Naryeong Kim, and Shin Yoo are with the School of Computing, KAIST, Daehak-ro 291, Daejeon, South Korea.\protect\\
E-mail: \{gabin.an, jinsuchoi, kimnal1234, shin.yoo\}@kaist.ac.kr.
\IEEEcompsocthanksitem Jingun Hong is with SAP Labs Korea, Seoul, South Korea.\protect\\
E-mail: jingun.hong@sap.com.}%
}

\markboth{}%
{An \MakeLowercase{\textit{et al.}}: Identifying Bug Inducing Commits by Combining Fault Localisation and Code Change Histories}


\maketitle

\begin{abstract}
  A Bug Inducing Commit (BIC) is a code change that introduces a bug into the codebase. Although the abnormal or unexpected behavior caused by the bug may not manifest immediately, it will eventually lead to program failures further down the line. When such a program failure is observed, identifying the relevant BIC can aid in the bug resolution process, because knowing the original intent and context behind the code change, as well as having a link to the author of that change, can facilitate bug triaging and debugging. However, existing BIC identification techniques have limitations. Bisection can be computationally expensive because it requires executing failing tests against previous versions of the codebase. Other techniques rely on the availability of specific post hoc artifacts, such as bug reports or bug fixes. In this paper, we propose a technique called \name that aims to identify the BIC with a core concept that a commit is more likely to be a BIC if it has more recently modified code elements that are highly suspicious of containing the bug. To realise this idea, \name leverages two fundamental relationships in software: the failure-to-code relationship, which can be quantified through fault localisation techniques, and the code-to-commit relationship, which can be obtained from version control systems. Our empirical evaluation using 206 real-world BICs from open-source Java projects shows that \name significantly outperforms state-of-the-art BIC identification techniques, achieving up to 45.8\% higher MRR. We also report that the ranking scores produced by \name can be used to perform weighted bisection. Finally, we apply \name to a large-scale industry project with over 10M lines of code, and show that it can rank the actual BIC within the top five commits for 87\% of the studied real batch-testing failures, and save the BIC inspection cost by 32\% on average.

\end{abstract}


\section{Introduction}
\IEEEPARstart{I}n modern software development workflows based on Continuous Integration/Continuous Deployment (CI/CD), numerous developers simultaneously participate in the development of a single project, and multiple code changes (or commits) are continuously integrated into a shared repository. In such an environment, when an abnormal program behavior (i.e., program failure) is observed during the testing process or when a field failure occurs after release, the QA or development teams analyse and categorise the issue, assign it to the most suitable developer, who then performs debugging activities (this process is referred to as the bug resolution process). During the software development process, the commits that contain buggy source code, which eventually leads to program failures, are known as Bug Inducing Commits (BICs)~\cite{liwerski2005}. For a particular program failure, knowing which commit in the change history is more likely to be a BIC can provide significant advantages in effectively performing the bug resolution process. Firstly, it can streamline the bug assignment phase by enabling the effective assignment of a newly discovered bug to the right team or developers. This is facilitated by the connection between commits and their respective authors, coupled with the fact that 78\% of bugs are ultimately fixed by the developers who originally introduced them~\cite{Wen2016}. Secondly, it can assist in both automated and manual debugging activities. Prior work has shown that simply reverting BICs may suffice for bug fixes~\cite{Wu2017, Wen2020}, and BIC information can be utilized to improve the accuracy of Fault Localisation (FL) techniques~\cite{Wen2021}. Furthermore, the knowledge of BICs has been demonstrated to aid developers in manual debugging efforts~\cite{Wen2019, Wen2021}. Finally, it can reduce testing costs, especially in batch testing failure scenarios where tests are executed against a cumulative batch of changes to reduce overall testing costs, but some tests fail~\cite{Beheshtian2022}. In such cases, knowing the relative suspiciousness of commits in the batch can expedite the identification of the specific commit that caused the failure, potentially reducing the number of test executions required compared to the standard bisection (we show this through our experiment in the later of this paper). 

Recognising the usefulness of identifying BICs for a given failure, multiple BIC identification techniques have been proposed, which can be broadly categorised into two distinct groups. The first group employs a conventional approach known as bisection~\cite{gitbisect}, which conducts a binary search over the commit history, systematically evaluating each past program snapshot to identify whether it manifests the buggy behavior or not. The evaluation process can be performed either manually or through an automated execution of the test cases that reveal the bug. However, even with automation, the bisection process can incur significant computational overhead if building and testing a specific program version is resource-intensive~\cite{Wen2021}.
The second group consists of Information Retrieval (IR)-based BIC identification techniques~\cite{Wen2016, Bhagwan2018}, also known as changeset/commit-level fault localisation. These approaches reformulate BIC identification as an IR problem. While various information about the failure in a textual format, such as a bug report, is treated as a query, the commits are considered as documents. The approaches then identify the BIC by directly assessing the lexical or semantic similarity between the bug report and the commits. Although IR-based approaches do not incur the computational costs associated with their dynamic counterparts (e.g., building and testing), they have limitations. These techniques can only leverage textual information, making it challenging to capture complex failure behaviors. Additionally, their effectiveness heavily relies on the quality and completeness of the bug reports. Other than these two groups, there is a family of BIC identification techniques represented by the SZZ algorithm~\cite{liwerski2005} and its variants, but they are not applicable during the debugging time as they require a Bug Fixing Commit (BFC) as input.

\begin{figure}[t]
  \centerline{\includegraphics[width=\linewidth]{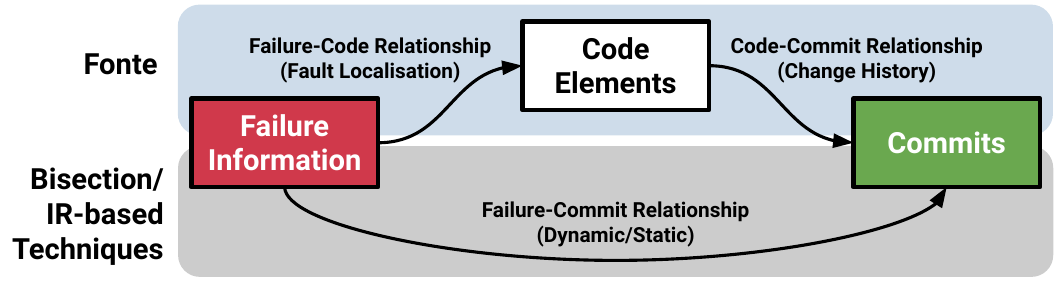}}
  \caption{A comparison of BIC identification methods, highlighting the distinctions between \name, Bisection and IR-based Techniques, and their relationships with failure information, code elements, and commits.}
  \label{fig:fonte-conceptual-comparison}
\end{figure}

In this paper, we introduce \name\footnote{\name is an Italian word meaning ``source'' or ``origin''.}, a novel unsupervised approach for identifying BICs that is efficient, flexible, and available during the debugging phase. As shown in \Cref{fig:fonte-conceptual-comparison}, unlike bisection and the IR-based BIC identification methods that attempt to directly assess the relevance between the commit and the failure (or bug report), \name first quantifies the relevance of code elements to the failure, computed by a FL technique~\cite{Wong2016}, and maps these code-level scores to the commit level using the change history of the code elements. This process extends existing FL techniques beyond their traditional spatial domain (i.e., the location within the code) to a temporal one (i.e., the commit in the history of the codebase).
Compared to the bisection approach, \name requires examining only the present buggy code version where the bug manifests, making it more efficient. Additionally, \name has the ability to incorporate a wide array of failure-specific information, from test coverage to bug reports, since it is compatible with any FL technique that yields quantitative suspiciousness scores, such as Spectrum-based Fault Localisation (SBFL) or Information-Retrieval-based Fault Localisation (IRFL). Consequently, we expect that \name can naturally benefit from future advancements in FL techniques, as its design allows for seamless integration of novel FL approaches.


We evaluate \name using a benchmark of 206 real-world Java bugs from Defects4J~\cite{Just2014} and their ground-truth BICs, comprising 67 from an existing BIC dataset~\cite{Wen2019}, and 139 that are manually curated by us. The results demonstrate that the ranking performance of \name, when combined with the Ochiai SBFL technique, significantly outperforms state-of-the-art IR-based BIC identification techniques, achieving up to 45.8\% higher Mean Reciprocal Rank (MRR). Furthermore, we propose a weighted bisection algorithm that leverages the commit scores produced by \name during the search process. Our findings show that weighted bisection can reduce the number of search iterations for 98\% of the cases compared to the standard bisection, respectively. Finally, we also apply \name to the historical CI data of \product, showcasing its practical applicability in an industry setting.

This paper presents an extended version of our previously published work~\cite{An2023}. The main extensions and contributions over the original work are summarised as follows:

\begin{itemize}
  \item \textbf{Evaluating Various Code History Tracking Tools}: While \name is compatible with any method-level code history tracking tool, our previous work only used the \texttt{git log} command to obtain the code-commit relationship. This extended study makes use of the recently proposed code change tracking tools, CodeShovel~\cite{Grund2021} and CodeTracker~\cite{Jodavi2022}, in addition to \texttt{git log} and reports their effectiveness regarding the search space reduction and recall, and also the efficiency of collecting code history.
  \item \textbf{Evaluating Diverse Failure-Code Relationships}: The previous work uses only SBFL to measure the failure-code relationships in the experiment. This extended study incorporates not only SBFL but also IRFL to establish the relationships between a failure and code elements when applying \name, and compare \name's performance based on the underlying FL technique. Furthermore, the inclusion of IRFL enables a fairer comparison between our methodology and other IR-based approaches for BIC identification.
  \item \textbf{More Realistic Assessment of FL Accuracy's Effect on \name's Performance}: Previously, to study the impact of FL accuracy on \name's effectiveness, SBFL's performance was artificially weakened by removing some tests from the test suite. In this extended work, since we incorporate both IRFL and SBFL with \name, we directly evaluate \name's performance in relation to the inherent accuracy of the underlying FL techniques. This approach provides a more authentic assessment without the need for artificial manipulations.
  \item \textbf{Richer Evaluation Dataset}: We have expanded the BIC dataset from the initial 130 bugs in our previous paper~\cite{An2023} to 206 bugs for a more comprehensive and diverse evaluation. This extended dataset contains bugs from all 16 Java projects that use Git as their version control system in the Defects4J 2.0 benchmark~\cite{Just2014}, whilst the previous dataset contained bugs from 11 projects. The reproducing package of \name with the new dataset is publicly available at our GitHub repository\footnote{\url{https://github.com/coinse/fonte/blob/extension}}.
\end{itemize}







The remainder of the paper is structured as follows. Section~\ref{sec:background} explains the research context of this paper and defines the basic notations. Section~\ref{sec:methodology} and \ref{sec:weighted_bisection} propose \name and the novel weighted bisection method, respectively. Section~\ref{sec:eval_setup} describes the empirical settings for \name along with the research questions, and Section~\ref{sec:results} presents the results. Section~\ref{sec:industry} shows the application results of \name to the batch testing scenario in industry software. Section~\ref{sec:threats} addresses the threats to validity, and Section~\ref{sec:related_work} covers the related work of \name. Finally, Section~\ref{sec:conclusion} concludes.

\section{Background}
\label{sec:background}

This section provides the background of this paper.

\subsection{Research Context}
The debugging process is typically initiated by observing a failure that reveals the presence of a bug in the software. Prior research indicates that a single failing test case is often the most commonly available information when debugging begins~\cite{Kochhar2016}. Even when users report a field failure, the debugging activities typically commence with reproducing the failure~\cite{Artzi, Jin2012,Zimmermann2010}. This is because failure-triggering test cases are essential for confirming whether the bug has been successfully fixed or not. Once a program failure is observed and reproduced, identifying the BIC responsible for the failure can contribute to a more efficient bug triage process~\cite{Murali2021} and aid developers in better understanding the context of the buggy behaviour~\cite{Wen2016}.

While some BIC identification techniques~\cite{Bhagwan2018,Murali2021} rely on information derived from failures, such as stack traces or exception messages (in a textual format), these sources may only be indirectly linked to the contents of actual BICs. Since commits are directly coupled to specific locations in the source code, our approach focuses on leveraging the actual coverage of the failing tests as the primary source of information. Our previous work~\cite{An2021} has shown that the coverage of failing test executions, referred to as \emph{failure coverage}, can effectively reduce the search space for BICs. By simply filtering out any commit that is not related to the evolution of code elements covered by the failing tests, the search spaces for 703 bugs in the Defects4J v2.0 benchmark~\cite{Just2014} were reduced to an average of 12.4\% of their original size. This significant reduction rate suggests that failure coverage has the potential to provide a solid foundation for a BIC identification technique available during the debugging phase.

The objective of this work is to accurately identify the BIC using solely the information available at the onset of debugging, immediately following the observation and reproduction of a test failure. Building upon our previous technique for reducing the BIC search space~\cite{An2021}, we introduce an approach that can precisely quantify the relevance of each commit within the reduced search space to the observed failure. Rather than directly measuring the relevance of commits to a failure, our method leverages two fundamental types of relationships, namely the failure-code relationship and the code-commit relationship, and combines these relationships to derive the failure-commit relationships.
\subsection{Basic Notations}
\label{sec:background:notation}

Let us define the following properties of a program $P$:
\begin{itemize}

  \item A set of commits $C = \{c_1, c_2, \ldots\}$ made to $P$
  \item A set of code elements $E = \{e_1, e_2, \ldots\}$ of $P$, such as
  statements or methods

  \item A set of test cases $T = \{t_1, t_2, \ldots\}$ where $T_F \subseteq T$ is a set of failing test cases
\end{itemize}

We assume that there is at least one failing test case, i.e., $|T_F| > 0$, and the bug responsible for the failure resides in the source code, i.e., some elements in $E$ cause the failure of $T_F$. We also define the following relations on sets $C$, $T$, and $E$:

\begin{itemize}
  \item A relation $\mathsf{Cover} \subseteq T \times E$ defines the relation between test cases and code elements in the program $P$. For every $t \in T$ and $e \in E$, $(t, e) \in \mathsf{Cover}$ if and only if the test $t$ covers $e$ during the execution.
  \item A relation $\mathsf{Evolve} \subseteq C \times E$ defines the relation between past commits and code elements in the program $P$. For every $c \in C$ and $e \in E$, $(c, e) \in \mathsf{Evolve}$ if and only if the commit $c$ is in the change history of the code element $e$.
\end{itemize}

As our ultimate goal is to find the BIC in $C$, we aim to design a scoring function $s \colon C \to \mathbb{R}$ that gives higher scores to commits that have a higher probability of being the BIC.

\section{\name: Automated BIC Identification via Dynamic, Syntactic, and Historical Analysis}
\label{sec:methodology}

This paper presents \name, a technique to automatically identify the BIC, based
on the assumption that \emph{a commit is more likely to be a bug inducing
commit if it introduced or modified a code element that is more relevant to
the observed failure}. The key idea behind \name is that the relevancy of the code elements to the observed failures can be quantified using existing FL techniques~\cite{Wong2016}, such as SBFL or IRFL, while the association of commits with code elements can be obtained by employing change history tracking tools, such as \texttt{git log}, CodeShovel~\cite{Grund2021}, or CodeTracker~\cite{Jodavi2022}. \Cref{fig:overview} illustrates the three stages of
\name, which are described below:

\begin{figure}[t]
  \centerline{\includegraphics[width=\linewidth]{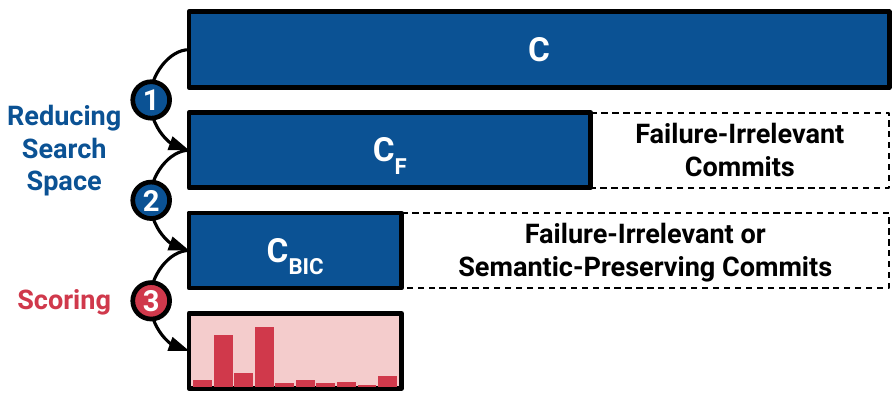}}
  \caption{The three-stage process of \name from the reduction of the BIC search space to the final commit scoring stage.}
  \label{fig:overview}
\end{figure}

\begin{enumerate}
\item \name identifies all potentially buggy code elements using the coverage of failing test cases and discards the commits that are irrelevant to those code elements~\cite{An2021}.
\item \name additionally filters out the semantic-preserving commits that contain only style changes to the suspicious files using AST-level comparisons.
\item \name computes suspiciousness scores of the remaining commits to rank the commits in terms of potential responsibility for the failure.
\end{enumerate}

The rest of this section describes each stage in more detail.

\subsection{Stage 1: Filtering Out Failure-Irrelevant Commits}
\label{sec:methodology:stage1}

Using the notations defined in Section~\ref{sec:background:notation}, we can represent the failure-coverage-based BIC search space reduction~\cite{An2021} as follows. First, let $E_{F} \subseteq E$ denote the set of all code elements that are covered by the failing test cases:

\begin{equation}
  \label{eq:E_susp}
E_{F} = \bigcup_{t \in T_F}\{e \in E|(t, e) \in \mathsf{Cover}\}
\end{equation}

Subsequently, we obtain $C_{F} \subseteq C$, a set of commits that are involved in the evolution of at least one code element in $E_{F}$:
\begin{equation}
  \label{eq:C_susp}
C_{F} = \bigcup_{e \in E_{F}}\{c \in C|(c, e) \in \mathsf{Evolve}\}
\end{equation}

Then, all commits not contained in $C_{F}$ can be discarded from our BIC
search space because the changes introduced by those commits are not related to
any code element executed by failing executions. Consequently, the BIC search
space is reduced from $C$ to $C_{F}$.

\subsection{Stage 2: Filtering Out Semantic-Preserving Commits}
\label{sec:methodology:stage2}

\begin{figure}[t]
\begin{lstlisting}[language=diff]
@@ -74,7 +74,7 @@ import org.apache.commons.lang.exception.NestableRuntimeException;
  * @author Phil Steitz
  * @author Pete Gieser
  * @since 2.0
- * @version $Id: StringEscapeUtils.java,v 1.26 2003/09/07 14:32:34 psteitz Exp $
+ * @version $Id: StringEscapeUtils.java,v 1.27 2003/09/13 03:23:24 psteitz Exp $
  */
 public class StringEscapeUtils {
 
@@ -242,7 +242,9 @@ public class StringEscapeUtils {
      } else {
          switch (ch) {
              case '\'':
-                 if (escapeSingleQuote) out.write('\\');
+                 if (escapeSingleQuote) {
+                     out.write('\\');
+                 }
                  out.write('\'');
                  break;
              case '"':
\end{lstlisting}
  \caption{Changes by the commit \texttt{5814f50} in Defects4J \texttt{Lang-46}}
  \label{fig:Lang-46b-5814f50}
\end{figure}

The reduced set of candidate BICs, $C_{F}$, may still contain \emph{semantic-preserving commits}, i.e., commits that do not introduce any semantic change to the suspicious code elements. These commits can be further excluded from the BIC search space, as they cannot have altered the functional behaviour of the program so thus cannot introduce a bug~\cite{Kim2006}. An example of such a commit is shown in \Cref{fig:Lang-46b-5814f50}, which modifies the comments and encloses the single statement in the \texttt{if} block.

We use the file-level AST level comparison~\cite{Falleri2014} to identify whether a given commit $c \in C_{F}$ is a semantic-preserving commit or not. First, we identify the set of failure-relevant source files, $S$, that are modified by the commit $c$ and covered by the failing test cases. Formally, any file in $S$ contains at least one code element in:
\begin{equation}
E_{F,c} = \{e \in E_{F}|(c, e) \in \mathsf{Evolve}\}
\label{eq:E_susp_c}
\end{equation}
Subsequently, for each file $s \in S$, we compare the ASTs derived from $s$ before and after the commit $c$.\footnote{The AST comparison information per file for each commit can be computed and stored promptly upon its creation, and if consistently updated in CI environments, there is no additional cost associated with computing it when applying \name.} If the ASTs are identical for all files in $S$, we consider the commit $c$ as a semantic-preserving commit.
Note that this approach does
not guarantee 100\% recall, as it is possible for two source files to yield
different ASTs while sharing the same semantic.
However, it can safely prune the search space due to its soundness, i.e., if it identifies a commit as semantic-preserving, it is guaranteed to be semantic-preserving. Consequently, the search space for BIC can be further reduced to $C_{BIC} = C_{F} \setminus C_{SP}$, in which $C_{SP}$ denotes all identified semantic-preserving commits in $C_{F}$.

It is worth noting that during this phase, there is also the option to consider discarding the refactoring changes~\cite{fowler2018refactoring} as well, in the same manner as RA-SZZ~\cite{Neto2018}. However, existing tools for detecting refactoring code, such as RefDiff~\cite{Silva2021}, do not guarantee 100\% precision, which introduces the risk of mistakenly excluding semantic-changing commits. Therefore, to guarantee the completeness of the narrowed-down BIC search space, \name only employs reliable AST-level comparisons.

\subsection{Stage 3: Scoring Commits using FL Scores and Code Change History}
\label{sec:methodology:stage3}

After reducing the search space, the remaining task is to evaluate how likely it is that each commit within $C_{BIC}$ is responsible for the observed failure. As mentioned earlier, our basic intuition is that if a commit had created, or modified, more suspicious code elements for the observed failures, it is more likely to be a BIC.

The suspiciousness of code elements can be quantified via an FL technique. For
example, we can apply SBFL~\cite{Wong2016} using the coverage of the test suite
$T$: note that SBFL uses only test coverage and result information, both of
which are available at the time of observing a test failure. Assuming that we
are given the suspiciousness scores, let $susp \colon E_{F} \to \mathbb{R}^{\geq 0}$ be the mapping function from each suspicious code element in $E_{F}$
to its non-negative FL score.\footnote{The constraint of FL-score being
non-negative is adopted for the sake of simplicity. Note that any FL results
can be easily transformed so that the lowest score is 0.} To convert the
code-level scores to the commit level, we propose a voting-based commit scoring
model where the FL score of a code element is distributed to its relevant
commits. The model has two main components: rank-based voting power and
depth-based decay.

\begin{table}[t]
  \centering
  \caption{Example of the voting power of code elements}
  \scalebox{1.00}{
  \begin{tabular}{l|l|rrrrr}
  \toprule
  \multicolumn{2}{l|}{\textbf{Code Element}}           & $e_1$ & $e_2$ & $e_3$ & $e_4$ & $e_5$\\\midrule
  \multicolumn{2}{l|}{\textbf{Score}}            & 1.0   & 0.6   & 0.6 & 0.6 & 0.3  \\\midrule
  \multicolumn{2}{l|}{$rank_{\mathtt{max}}$}  & 1     & 4    & 4     & 4 & 5\\
  \multicolumn{2}{l|}{$rank_{\mathtt{dense}}$}  & 1     & 2    & 2     & 2 & 3\\\midrule
  \multirow{4}{*}{$vote$}&$\alpha=0$, $\tau=\mathtt{max}$   & 1.00  & 0.25  & 0.25  & 0.25 & 0.20  \\
  &$\alpha=1$, $\tau=\mathtt{max}$   & 1.00  & 0.15  & 0.15  & 0.15  & 0.06  \\
  &$\alpha=0$, $\tau=\mathtt{dense}$ & 1.00  & 0.50  & 0.50  & 0.50  & 0.33  \\
  &$\alpha=1$, $\tau=\mathtt{dense}$ & 1.00  & 0.30  & 0.30  & 0.30  & 0.10  \\
\bottomrule
  \end{tabular}}
  \label{tab:voting}
\end{table}

\textbf{Rank-based Voting Power:} Recent work~\cite{Sohn2019,Sohn2021,
habchi2022made} showed that, when aggregating FL scores from finer granularity
elements (e.g, statements) to a coarser level (e.g., methods), it is better to
use the \emph{relative rankings} from the original level only, rather than directly
using the raw scores. The actual aggregation takes the form of voting: the higher
the ranking of a code element is in the original level, the more votes it is
assigned with for the target level. Subsequently, each code element casts its
votes to the related elements in the target level. We adopt this voting-based
method to aggregate the statement-level FL scores to commits. The
\emph{voting power} of each code element $e$ based on their FL rankings (and scores) as follows:

\begin{equation}
vote(e) = \frac{\alpha*susp(e) + (1 - \alpha)*1}{rank_{\tau}(e)}
\label{eq:vote}
\end{equation}

where $\alpha \in \{0, 1\}$ is a hyperparameter that decides whether to use the
suspiciousness value ($\alpha=1$) as a numerator or not ($\alpha=0$), and $\tau$
a hyperparameter that defines the tie-breaking scheme. We vary $\tau \in \{\mathtt{max},
\mathtt{dense}\}$: the $\mathtt{max}$ tie-breaking scheme gives the lowest (worst)
rank in the tied group to all tied elements, while $\mathtt{dense}$ gives the highest
but does not skip any ranks after ties. By design, $\tau=\mathtt{max}$ will penalise tied
elements more severely than $\tau=\mathtt{dense}$.
The example in \Cref{tab:voting} shows how the hyperparameters affect voting. Note that the relative order between FL scores is preserved in the voting power regardless of hyperparameters, i.e., $vote(e) > vote(e')$ if and only if $susp(e) > susp(e')$.

\textbf{Depth-based Decay:}
Wen et al.~\cite{Wen2016} showed that using the information about commit time
can boost the accuracy of the BIC identification. Similarly, Wu et al.~\cite{Wu2017} observed that the commit time of crash-inducing changes is closer to
the reporting time of the crashes.
Based on those findings, we hypothesise that \emph{older commits are less likely to
be responsible for the currently observed failure}, because if an older commit was
a BIC, it is more likely that the resulting bug has already been found and
fixed.
To capture this intuition, our approach diminishes the voting impact of a program element for older commits that have greater historical depths. The historical depth of a commit $c$, with respect to a code element $e \in E_{F,c}$ (\Cref{eq:E_susp_c}), is defined as follows:
\begin{equation}
\begin{aligned}
depth(e,c) = & |\{c' \in C_{BIC}|\\
& (c', e) \in \mathsf{Evolve} \wedge c'.time > c.time\}|
\end{aligned}
\label{eq:depth}
\end{equation}
which is essentially a count of the commits in the set $C_{BIC}$ that are relevant to a particular code element $e$ and are more recent than the commit $c$.

\begin{figure}[t]
    \centering
    \includegraphics[width=0.95\linewidth]{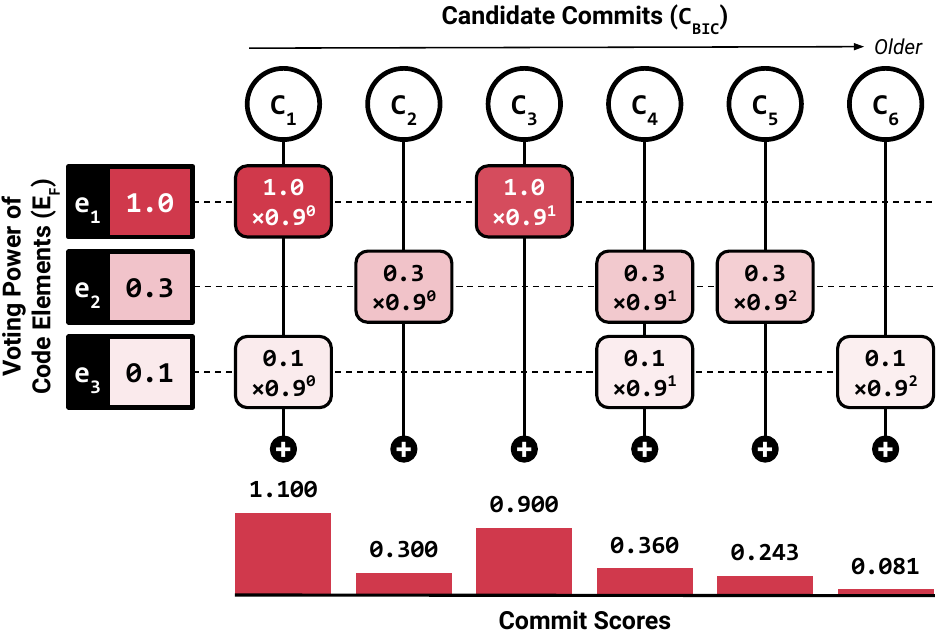}
    \caption{Example of computing the commit scores when $\lambda = 0.1$}
    \label{fig:voting}
  \end{figure}

Bringing it all together, we use the following model to assign a score to each commit $c$ in $C_{BIC}$:
\begin{equation}
  commitScore(c) = \sum_{e \in E_{F}^c}vote(e)*(1-\lambda)^{depth(e, c)}
\label{eq:commit_score}
\end{equation}
where $\lambda \in \left[0, 1\right)$ is the decay factor: when $\lambda=0$, there is no penalty for older commits. \Cref{fig:voting} shows the example of calculating the score of commits when $\lambda=0.1$.

Finally, based on $commitScore$, the commit scoring function $s \colon C \to
\mathbb{R}^{\geq 0}$ of \name is defined as follows:
$$s(c) =
  \begin{cases}
      commitScore(c)  & \text{if } c \in C_{BIC}\\
      0              & \text{otherwise}
  \end{cases}
$$

\section{Weighted Bisection}
\label{sec:weighted_bisection}

Bisection is a traditional way of finding the BIC by repeatedly narrowing down
the search space in half using binary search: it is also equipped in popular
Version Control Systems (VSCs), e.g., \texttt{git bisect} or \texttt{svn-bisect}. A standard 
bisection is performed as follows: given the last \emph{good} and earliest \emph{bad}
versions of a program, it iteratively checks whether the midpoint of
those two versions, referred to as a \emph{pivot}, contains the bug. If there
is a bug, the earliest bad point is updated to the pivot, otherwise, the last
good point is updated to the pivot. If there is a bug-revealing test case that
can automatically check the existence of a bug, this search process can be fully
automated.

However, as pointed out in previous work~\cite{Murali2021}, even though the bug
existence check can be automated, each bisect iteration may still require a
significant amount of time and computing resources, especially when the program
is large and complex, or the bug-revealing test takes a long time to execute.
Since a lengthy bisection process can block the entire debugging pipeline, we
aim to explore whether the bisection can be accelerated using the commit score
information.

\begin{figure}[t]
  \centerline{\includegraphics[width=0.95\linewidth]{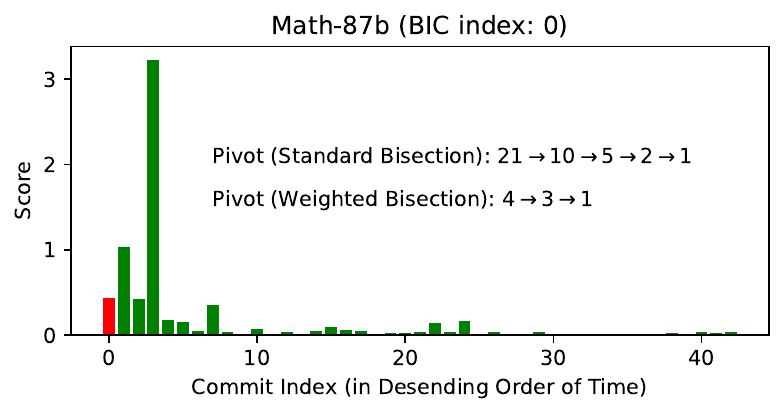}}
  \caption{Example of applying the weighted bisection to \texttt{Math-87}}
  \label{fig:search}
\end{figure}

We propose a \emph{weighted} bisection algorithm, where the search pivot is
set to a commit that will halve \emph{the amount of remaining commit scores}
instead of \emph{the number of remaining commits}, in order to to check
the point of greatest ambiguity where the chances of finding the BIC in
two directions (before and after the commit) are almost equal. For example, let us consider
the example in \Cref{fig:search} that shows the score distribution of the
commits in the reduced BIC search space of \texttt{Math-87} in Defects4J. For
\texttt{Math-87}, the score distribution is biased towards a small number of
recent commits including the real BIC (marked in red) with the third-highest score. In this
case, simply using the midpoint as a search pivot might not be a good choice
because all highly suspicious commits still remain together on one side of the
split search space: as a result, the standard bisection requires five iterations to finish.
Alternatively, if we pivot at the commit that halves the amount of
remaining scores, the bisection reaches the actual BIC more quickly, completing
the search in three iterations.

\begin{algorithm}[t]
  \small
  \SetCommentSty{mycommfont}
  \SetKwInput{KwPrecondition}{Precondition}
  \SetKwInput{KwPostcondition}{Postcondition}
  \SetKwFunction{CScore}{s}

  \KwIn{Array of commits $C$}
  \KwIn{Commit score (weight) function $s: C \rightarrow \mathbb{R}^{\geq 0}$}
  \KwOut{Bug inducing commit $c \in C$}
  \KwPrecondition{$C[i]$ is newer than $C[j]$ if and only if $i < j$}
  \KwPostcondition{$\mathtt{bad} + 1 = \mathtt{good}$}

  $C' \longleftarrow$ empty list\;
  \For{$c \in C$}{
      \If{$s(c) > 0$}{ \tcp{Extract Subarray with Positive Scores}
          Append $c$ to $C'$\;
      }
  }
  $\mathtt{bad}, \mathtt{good} \longleftarrow 0, |C'|$\\
  \While{$\mathtt{bad} + 1 < \mathtt{good}$}{
    $\mathtt{pivot} \leftarrow \text{argmin}_{p=\mathtt{bad}+1}^{\mathtt{good}-1}{|\sum_{i=\mathtt{bad}}^{p-1}s(C'[i]) - \sum_{i=p}^{\mathtt{good}-1}s(C'[i])|}$\\
    \If{The target bug is detected in $C'[\mathtt{pivot}]$}{
      $\mathtt{bad} \leftarrow \mathtt{pivot}$\\
    }\Else{
      $\mathtt{good} \leftarrow \mathtt{pivot}$\\
    }
  }

  \Return{$C'[\mathtt{bad}]$}
  \caption{Weighted Bisection Algorithm}
  \label{algo:search}
\end{algorithm}

Algorithm~\ref{algo:search} presents the weighted bisection algorithm.
It takes as input a chronologically sorted array of commits $C$ and a commit score function $s \in C \rightarrow \mathbb{R}^{\geq 0}$, which assigns non-negative scores to each commit, and returns the BIC.
First, it extracts a subarray $C'$ from $C$, containing only the commits with positive scores (Lines 1-4). Assuming that at least one BIC exists in the sorted sequence $C'$, the earliest bad index $\mathtt{bad}$ is initialized to 0, representing the index of the most recent commit (Line 5). Since all commits in $C'$ are BIC candidates, the last good index $\mathtt{good}$ is set to the (virtual) index immediately after the oldest commit (Line 5). The algorithm then iteratively selects a new $\mathtt{pivot}$ index from the range $[\mathtt{bad}+1, \mathtt{good}-1]$, until there are no remaining commits between $\mathtt{bad}$ and $\mathtt{good}$ (Line 6). The $\mathtt{pivot}$ selection is performed such that it minimizes the difference between the sum of scores on the left side (not including $\mathtt{pivot}$) and the sum of scores on the right side (including $\mathtt{pivot}$) (Line 7). Once a new $\mathtt{pivot}$ is selected, the commit $C'[\mathtt{pivot}]$ is inspected for the bug, either by executing the bug-revealing tests or through manual inspection (Line 8). If the bug is detected (i.e., the $\mathtt{pivot}$ is a bad commit), the $\mathtt{bad}$ index is updated to $\mathtt{pivot}$ (Line 9); otherwise (i.e., the $\mathtt{pivot}$ is a good commit), the $\mathtt{good}$ index is updated to $\mathtt{pivot}$ (Line 11). Finally, when the loop terminates, the algorithm returns the identified BIC at the $\mathtt{bad}$ index (Line 12).

It is worth noting that this algorithm is a \emph{generalised} version of the standard bisection: the standard bisection method can be considered a special case of the weighted bisection algorithm where $s$ is a non-zero constant function, assigning the same non-zero value to all commits, e.g., $\forall c \in C: s(c) = 1.0$.

\section{Evaluation Setup}
\label{sec:eval_setup}

This section describes the dataset used for evaluation (\Cref{sec:dataset}), the details of \name's implementation (\Cref{sec:implementation}), and our research questions along with their experimental protocols (\Cref{sec:rq}).

\subsection{Dataset of Bug Inducing Commits}
\label{sec:dataset}
We choose Defects4J v2.0.0~\cite{Just2014}, a collection of 835 real-world bugs
in Java open-source programs, as the source of our experimental subjects.
While Defects4J provides test suites containing the bug-revealing tests
for every bug, as well as the entire commit history for each buggy version, it
lacks the BIC information for each bug.

We, therefore, start with a
readily-available BIC dataset for 91 Defects4J bugs\footnote{https://github.com/justinwm/InduceBenchmark} constructed by Wen et al.~\cite{Wen2019}. This
dataset was created by running the bug-revealing test cases on the past
versions and finding the earliest buggy version that makes the tests fail.
However, in our experiment, we are forced to exclude 24 out of 91 data points.
Since \name is implemented using Git, it cannot trace the commit history of
nine bugs from the \texttt{JFreeChart} project which uses SVN as its version
control system. Further, we exclude 14 data points that are shown to be
inaccurate by previous work~\cite{An2021}. Lastly, \texttt{Time-23} is also
discarded, because we found that the identified commit in the dataset does not
contain any change to code, but only to the license comments. The detailed
reasons can be found in our repository. In summary, we make use of 67
ground-truth BICs from this dataset.

The original dataset of 67 ground-truth BICs identified by Wen et al.~\cite{Wen2019} encompasses bugs from only four out of the 17 projects in Defects4J. To expand the evaluation dataset, we have manually compiled an additional set of ground-truth BICs. Two authors independently pinpointed the BIC for each bug by examining bug reports, symptoms of failures, and patches provided by developers. To minimize the effort required for manual inspection, our initial focus was on all Defects4J bugs where the narrowed down BIC search space, $C_{BIC}$, comprised ten or fewer potential BICs. This process led to a consensus on 70 instances, which were then included in the dataset. Furthermore, we identified ground-truth BICs for more bugs, mainly from software projects that initially had minimal or no BIC data. An additional 76 data points, on which the authors agreed, were also incorporated. In summary, a total of 206 data points (67 from Wen et al. + 146
manually curated - 7 overlapped) are used for the evaluation of \name. \Cref{tab:subjects} shows the number of ground-truth BICs for each software project, before and after the manual curation of the additional dataset compared to the initial dataset from Wen et al.~\cite{Wen2019}. The
combined BIC dataset and the provenance of each data point are available in our
repository for further scrutiny.

\begin{table}[t]
    \centering
    \caption{Distribution of ground-truth BICs across software projects: a comparison between the initial dataset taken from Wen et al.~\cite{Wen2019} and the augmented dataset after manual data curation.}
    \scalebox{1.00}{
    \begin{tabular}{l|rr}
        \toprule
        Project & \multicolumn{2}{c}{\# BICs}\\\cmidrule{2-3}
         & Wen et al.~\cite{Wen2019} & Augmented\\\midrule
        Cli & 0  & 15\\
        Closure & 35 & 36\\
        Codec & 0 & 3\\
        Collections & 0 & 2\\
        Compress & 0 & 5\\
        Csv & 0 & 13\\
        Gson & 0 & 4\\
        JacksonCore & 0 & 5\\
        JacksonDatabind & 0 & 6 \\
        JacksonXml & 0 & 2\\
        Jsoup & 0 & 35\\
        JxPath & 0 & 4\\
        Lang & 6 & 29\\
        Math & 21 & 40\\
        Mockito & 0 & 1\\
        Time & 5 & 6\\\midrule
        Total & 67 & 206\\\bottomrule
    \end{tabular}}
    \label{tab:subjects}
\end{table}

\subsection{Implementation Details of \name}
\label{sec:implementation}

\begin{table}[t]
  \centering
  \caption{Example of Relevant Test Selection (\texttt{Time-15})}
  \scalebox{0.95}{
  \begin{tabular}{l}
  \toprule
  \textbf{Failing Test ($T_F$)}\\\midrule
  org.joda.time.field.TestFieldUtils::testSafeMultiplyLongInt\\\midrule
  \textbf{Classes Covered by the Failing Test}\\\midrule
  org.joda.time.field.\textbf{FieldUtils}\\
  org.joda.time.\textbf{IllegalFieldValueException}\\\midrule
  \textbf{Relevant Tests ($T \setminus T_F$)}\\\midrule
  org.joda.time.Test\textbf{IllegalFieldValueException}::testGJCutover\\
  org.joda.time.Test\textbf{IllegalFieldValueException}::testJulianYearZero\\
  org.joda.time.Test\textbf{IllegalFieldValueException}::testOtherConstructors\\
  org.joda.time.Test\textbf{IllegalFieldValueException}::testReadablePartialValidate\\
  org.joda.time.Test\textbf{IllegalFieldValueException}::testSetText\\
  org.joda.time.Test\textbf{IllegalFieldValueException}::testSkipDateTimeField\\
  org.joda.time.Test\textbf{IllegalFieldValueException}::testVerifyValueBounds\\
  org.joda.time.Test\textbf{IllegalFieldValueException}::testZoneTransition\\
  org.joda.time.field.Test\textbf{FieldUtils}::testSafeAddInt\\
  org.joda.time.field.Test\textbf{FieldUtils}::testSafeAddLong\\
  org.joda.time.field.Test\textbf{FieldUtils}::testSafeMultiplyLongLong\\
  org.joda.time.field.Test\textbf{FieldUtils}::testSafeSubtractLong\\
  \bottomrule
  \end{tabular}}
  \label{tab:relevant}
\end{table}

\subsubsection{Basic Properties}
We apply \name at the \emph{statement}-level granularity. Following the notations in \Cref{sec:background:notation}, $E$ is a set
of statements composing the target buggy program. 
The initial BIC search space,
$C$, is set to all commits from the very first commit up to the commit
corresponds to the buggy version\footnote{\texttt{revision.id.buggy} in
Defects4J}. From the developer-written test cases, we only use the bug-revealing (i.e.,
failing) test cases as well as their relevant test cases as $T$. A test case is
considered relevant if and only if its full name contains the name of at least one class
executed by the failing test cases. \Cref{tab:relevant} shows the example
of the relevant test selection.

\subsubsection{Construction of the $\mathsf{Cover}$ relation}
To construct the $\mathsf{Cover}$ relation between $T$ and $E$, we measure the
statement-level coverage of each test case in $T$ using \texttt{Cobertura v2.0.3} which
is included in Defects4J.

\subsubsection{Construction of the $\mathsf{Evolve}$ relation}
To establish the $\mathsf{Evolve}$ relationship between $C$ and $E$, it is necessary to track the commit history for each code element. For this purpose, we utilise the \texttt{git log} command\footnote{\texttt{git log -C -M -L<start\_line>,<end\_line>:<file>}. The flags \texttt{-C} and \texttt{-M} are used for detecting file renaming, copying, or moving across versions.} in line with our prior research~\cite{An2021}. In addition to \texttt{git log}, we explored using CodeShovel~\cite{Grund2021} and CodeTracker~\cite{Jodavi2022}, which are advanced tools for retrieving the history of method changes. 

Please note that for each statement, we retrieve the commit history of its
enclosing method and create the $\mathsf{Evolve}$ relations between the
statements and the retrieved commits to ensure high recall for commit
histories. This is also to deal with omission bugs~\cite{Zhang2007}: if a bug is caused by
omission of some statements, we cannot trace the log of the missing statements
because they literally do not exist in the current version. In that case,
tracing the log of the neighbouring statements (in the enclosing method) will
enable to find the inducing commit, as the method that encloses the omission
bug should have been covered by the failing tests~\cite{An2021}.


\subsubsection{Detection of Semantic-Preserving Commits}
In Stage 2, before comparing the ASTs before and after a commit, we use OpenRewrite v7.21.0\footnote{https://github.com/openrewrite/rewrite} to ensure the same coding standard between the two versions of files. More
specifically, we use the \texttt{Cleanup} recipe\footnote{https://docs.openrewrite.org/reference/recipes/java/cleanup} that fixes any errors that violate CheckStyle rules.\footnote{https://checkstyle.sourceforge.io/}
This ensures that trivial differences between two versions that do not lead to
semantic differences are ignored: a good example is a commit in \texttt{Lang},
which is shown in \Cref{fig:Lang-46b-5814f50}. To compare ASTs after formatting the files, we use the
isomorphism test of GumTree v3.0.0~\cite{Falleri2014} that has time
complexity of $O(1)$.

\subsubsection{Fault Localisation\label{sec:fl}} In theory, any FL technique that produces suspiciousness scores (or rankings) can be plugged into \name. In this paper, we use two representative FL techniques, SBFL and IRFL.

\begin{itemize}
  \item SBFL: We use a widely-used SBFL formula, Ochiai~\cite{Abreu2006}, which can be expressed in our context as follows:
  $$
  Ochiai(e) = \frac{|\{t \in T_F|(t, e) \in \mathsf{Cover}\}|}{\sqrt{|T_F|*|\{t \in T|(t, e) \in \mathsf{Cover}\}|}}
  $$
  By definition, $Ochiai(e) > 0$ if and only if $e \in E_{F}$ (\Cref{eq:E_susp}).
  \item IRFL: We employ an unsupervised statement-level IRFL technique, Blues, which was proposed in a recent APR study~\cite{Motwani2023} and builds upon on BLUiR~\cite{Saha2013}, an unsupervised file-level IRFL technique. We directly utilise the pre-computed Blues suspiciousness scores for the Defects4J bugs available in their replication package.
\end{itemize}

\subsubsection{Hyperparameters}
\label{sec:hyperparameters}

In our experiments, we investigate the effects of varying hyperparameters: $\alpha \in \{0, 1\}$ and $\tau \in \{\mathtt{max},\mathtt{dense}\}$ for \Cref{eq:vote}, and $\lambda \in \{0.1, 0.2, 0.3\}$  for \Cref{eq:commit_score}.

\subsection{Research Questions}
\label{sec:rq}

We ask the following research questions in this paper:
\begin{itemize}
  \item \textbf{RQ1.} How accurately does \name find the BIC?
  \item \textbf{RQ2.} Does \name outperform other BIC identification approaches?
  \item \textbf{RQ3.} How efficient is the weighted bisection compared to the standard bisection?

\end{itemize}

\section{Results}
\label{sec:results}
This section describes the evaluation methodology used to investigate each research question, along with the findings.

\subsection{\textbf{RQ1. Effectiveness of \name}}

To evaluate the efficacy of \name, we pose three subsidiary questions focusing on its search space reduction capabilities, its performance in ranking BICs, and the impact of the ranking-based voting scheme and depth-based decay.

\begin{tcolorbox}[title=RQ1-1. Search Space Reduction (Stage 1\&2)]
To what extent do the first two stages of \name reduce the search space?
\end{tcolorbox}


\begin{table}[t]
\caption{Evaluation of code history tracking tools, focusing on the average reduction in BIC search space size, the validity ratio, and the average time taken to compile commit histories. The Validity Ratio represents the proportion of bugs for which the actual BIC is located within the reduced search space $C_{BIC}$.}
\label{tab:RQ1-reduction}
\scalebox{0.95}{
\begin{tabular}{r|rrr}
  \toprule
 History & Reduction Ratio & Validity Ratio & Duration\\
 Tracking Tool & (After Stage 1, Stage 2) & & \\\midrule
\texttt{git log} & 11.3\%, 11.0\% & $\mathbf{100\%}$ $(=\frac{206}{206})$ & \textbf{27.3s}\\
CodeShovel & 11.8\%, 11.1\% & $99.0\%$ $(=\frac{204}{206})$ & 521.0s\\
CodeTracker & \textbf{11.1\%}, \textbf{10.9\%} & $98.5\%$ $(=\frac{203}{206})$ & 738.3s\\\bottomrule
\end{tabular}}
\end{table}

\noindent\textbf{Evaluation Protocol} To determine the extent to which Stages 1 and 2 can narrow down the BIC search space, we calculate the reduction ratio of the BIC search space after Stage 1 (\(|C_{susp}|/|C|\)), and Stage 2 (\(|C_{BIC}|/|C|\)). Under a consistent methodology for measuring test coverage failures, the process of reducing the search space is influenced solely by the choice of code history tracking tools, which establish the \texttt{Evolve} (or code-commit) relation. Consequently, we report the outcomes of search space reduction using \texttt{git log}, CodeShovel~\cite{Grund2021}, and CodeTracker~\cite{Jodavi2022}. Additionally, we verify the presence of the actual BIC within the narrowed search space, \(|C_{BIC}|\), as a basic validity test, and measure the time taken to collect the commit histories using each tool to access the efficiency.

\noindent\textbf{Results} \Cref{tab:RQ1-reduction} presents a comparative analysis of commit counts filtered through the initial two stages of \name, using different code history tracking tools. \texttt{git log} and CodeTracker are the most efficient, reducing the commit count to 11.3\% and 11.1\% after the first stage, respectively, and further down to 11.0\% and 10.9\% after the second stage. CodeShovel shows a slightly less efficient reduction, with 11.8\% of commits remaining after the first stage and 11.1\% after the second. 
The Validity Ratio is highest for \texttt{git log} at 100\%. This means \texttt{git log} successfully retained the actual BIC in all evaluated instances. In contrast, CodeShovel and CodeTracker have slightly lower Validity Ratios at 99.0\% and 98.5\%, respectively, indicating that these tools occasionally produce incorrect commit histories, which is detailed in our artifect, which affects their reliability in retaining the BIC throughout the reduction process.
The Duration it takes for each tool to compile the commit histories is also noted, with \texttt{git log} being the fastest at 27.3 seconds on average, followed by CodeShovel at 521.0 seconds, and CodeTracker being the slowest at 738.3 seconds. This evaluation highlights the trade-offs between accuracy, efficiency, and effectiveness in reducing the BIC search space among the different tools.

\noindent\textbf{Answer to RQ1-1} On average, the BIC search space was narrowed down to approximately 11\% of its initial size, indicating that the choice of code history tracking tool had only a minimal impact on the reduction ratio. Among the evaluated tools, \texttt{git log} stands out for its reliability and efficiency.  

\begin{tcolorbox}[title=RQ1-2. Ranking Performance (Stage 3)]
  How effectively does the third stage of \name identify the BIC within the narrowed search space?
\end{tcolorbox}

\noindent \textbf{Evaluation Protocol} If the scoring mechanism of \name is effective, BICs will receive \emph{higher} scores compared to those changes that are not responsible for bugs. Therefore, to assess the performance of \name, we employ two widely recognized ranking-based metrics.
\begin{itemize}
  \item Mean Reciprocal Rank (MRR)~\cite{Craswell2009}: The average reciprocal rank of the BIC (\emph{higher is better})
  \item Accuracy@n (Acc@n): The percentage of bugs where the ranking of the ground-truth BIC is within the top $n$ positions (\emph{higher is better})
 \end{itemize} 
In situations where multiple elements share identical scores, we apply the max-tiebreaker approach, which conservatively allocates the worst rankings to these tied elements.
Additionally, we compare the performance of \name against a \textit{random baseline} to ascertain whether the ranking algorithm performs significantly better than random chance. In a scenario where $n$ commits exist in the reduced search space, the expected rank of a BIC under random conditions would be $\frac{1 + n}{2}$. We report the MRR and Acc@n values for this baseline.

Since \texttt{git log} demonstrated both reliable and efficient performance in RQ1-1, we use it to establish the \texttt{Evolve} relation for the remaining experiments. Additionally, we evaluate \name with diverse settings to answer this question. As a base FL technique for \name, we employ either SBFL or IRFL, as described in Section \ref{sec:fl}, in combination with diverse hyperparameter configurations outlined in Section \ref{sec:hyperparameters}.


\begin{table}
  \caption{The BIC ranking performance of \name combined with IRFL and SBFL, across various hyperparameter configurations}
  \label{tab:fonte_results}
  \scalebox{0.90}{
  \begin{tabular}{c|c|c|rrr|rrr}
    \toprule
    \multicolumn{3}{c|}{Hyperparameters}& \multicolumn{3}{c|}{\name with IRFL} & \multicolumn{3}{c}{\name with SBFL} \\\midrule
    $\lambda$ & $\alpha$ & $\tau$  & MRR & Acc@1 & Acc@5 & MRR & Acc@1 & Acc@5 \\
    \midrule
      \multirow{4}{*}{0.1} & \multirow{2}{*}{0} & $\mathtt{dense}$ & \textbf{0.453} & \textbf{30.6\%} & 62.6\% & 0.474 & 31.6\% & 67.5\% \\
                                               &  & $\mathtt{max}$ & 0.445 & 28.6\% & \textbf{65.0\%} & \textbf{0.481} & \textbf{32.0\%} & 68.9\% \\\
                           & \multirow{2}{*}{1} & $\mathtt{dense}$ & 0.433 & 27.2\% & 63.6\% & 0.475 & 30.6\% & \textbf{70.4\%} \\
                                               &  & $\mathtt{max}$ & 0.432 & 27.2\% & 63.1\% & 0.473 & 31.6\% & 69.9\% \\\midrule
      \multirow{4}{*}{0.2} & \multirow{2}{*}{0} & $\mathtt{dense}$ & 0.445 & 29.1\% & 61.7\% & 0.460 & 28.6\% & 68.4\% \\
                                               &  & $\mathtt{max}$ & 0.446 & 28.6\% & 64.6\% & 0.475 & 30.6\% & 69.4\% \\
                           & \multirow{2}{*}{1} & $\mathtt{dense}$ & 0.434 & 27.7\% & 62.6\% & 0.465 & 29.1\% & \textbf{70.4\%} \\
                                               &  & $\mathtt{max}$ & 0.436 & 28.2\% & 62.6\% & 0.465 & 30.1\% & 69.4\% \\\midrule
      \multirow{4}{*}{0.3} & \multirow{2}{*}{0} & $\mathtt{dense}$ & 0.445 & 29.1\% & 61.7\% & 0.451 & 27.7\% & 68.0\% \\
                                               &  & $\mathtt{max}$ & 0.434 & 26.7\% & 62.6\% & 0.471 & 30.6\% & 69.4\% \\
                           & \multirow{2}{*}{1} & $\mathtt{dense}$ & 0.421 & 25.2\% & 62.1\% & 0.458 & 28.2\% & 69.4\% \\
                                               &  & $\mathtt{max}$ & 0.422 & 25.2\% & 62.1\% & 0.459 & 28.6\% & 68.9\% \\\midrule
    \multicolumn{3}{c|}{Random Baseline} & 0.150 & 1.5\% & 32.5\% & 0.150 & 1.5\% & 32.5\%\\
    \bottomrule
    \end{tabular}}
\end{table}

\noindent \textbf{Results} 
\Cref{tab:fonte_results} details the effectiveness of \name in ranking BICs when integrated with two distinct FL techniques: IRFL and SBFL. The evaluation spans a range of hyperparameters and is quantified using three metrics: MRR, Acc@1, and Acc@5. The top-performing results for each FL technique and metric are denoted in bold within the table.

Generally, \name, regardless of the settings used, consistently exceeds the performance of the random baseline, demonstrating its strong capability to accurately identify the actual BIC among the potential commits in the reduced search space.
With IRFL, \name achieves its best MRR at 0.453 and its highest Acc@1 at 30.6\%. The best Acc@5 is 65.0\%. In the SBFL configuration, \name attains its best MRR of 0.481, an Acc@1 of 32.0\%, and an Acc@5 of 70.4\%. The hyperparameter setting of $\lambda=0.1$ consistently delivers superior results compared to other $\lambda$ values. Examining the voting hyperparameters $alpha$ and $tau$, for IRFL, the setting of $\alpha=0$ and $\tau=\mathtt{dense}$ demonstrates strong MRR and Acc@1 scores, while $\alpha=0$ and $\tau=\mathtt{max}$ is notable for its Acc@5. Conversely, for SBFL, the combination of $\alpha=0$ and $\tau=\mathtt{max}$ is notable for higher MRR and Acc@1, whereas $\alpha=1$ and $\tau=\mathtt{dense}$ is distinguished by its Acc@5 results.

\begin{figure}[t]
  \centerline{\includegraphics[width=\linewidth]{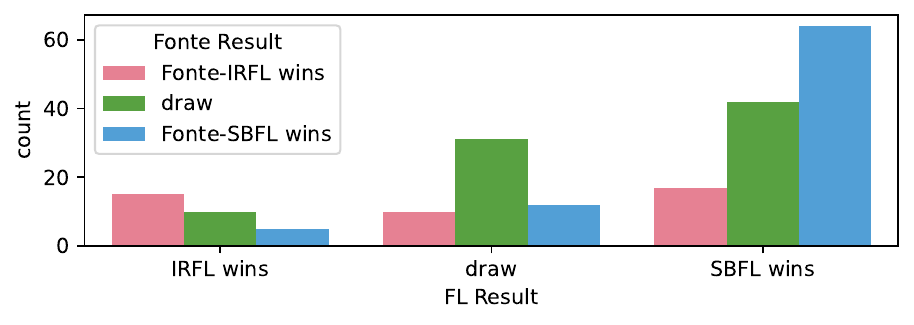}}
  \caption{Performance comparison of \name when integrated with IRFL and SBFL, categorised based on whether IRFL outperforms SBFL (IRFL wins), they have equal performance (draw), or SBFL outperforms IRFL (SBFL wins)}
  \label{fig:sbfl-irfl}
\end{figure}

We also observe that the integration of \name with SBFL yields superior results compared to combining it with IRFL. This advantage can be largely attributed to the higher FL accuracy of SBFL (using the Ochiai method) over IRFL (using the Blues method). When examining the average rankings of buggy methods across the analyzed bugs, IRFL outperforms SBFL in only 14.6\% (30 out of 206) of the cases, whereas SBFL surpasses IRFL in 60.0\% (123 out of 206) of the cases. \Cref{fig:sbfl-irfl} demonstrates that the performance of \name combined with IRFL outperforms that combined with SBFL in cases where IRFL outperforms SBFL, and vice versa. This overall trend indicates that more accurate FL results are likely to lead to better accuracy of \name in ranking BIC. Therefore, it is anticipated that \name will further benefit from advancements in more accurate and advanced FL techniques in the future.

\noindent\textbf{Answer to RQ1-2}: \name demonstrates high effectiveness in identifying the BIC within the narrowed search space. When combined with SBFL, \name achieves an MRR of 0.481, an Acc@1 of 32.0\%, and an Acc@5 of 70.4\%, indicating its ability to rank the actual BIC highly within the candidates.

\begin{tcolorbox}[title=RQ1-3. Ablation Study]
  What is the impact of Rank-based Voting Power and Depth-based Decay on the performance of \name?
\end{tcolorbox}

\noindent \textbf{Evaluation Protocol} To examine the influence of two key features of \name, Rank-based Voting Power and Depth-based Decay, on its overall performance, we conduct an ablation study. First, for evaluating the effect of Rank-based Voting Power, we evaluate a naive alternative strategy: using the FL scores directly as the voting power of each statement instead of considering the relative ranking among statements. In simpler terms, we replace \Cref{eq:vote} with the following equation:
\begin{equation}
\label{eq:vote_score}
vote(e) = susp(e)
\end{equation}
Second, to gauge the impact of Depth-based Decay, we measure how the performance of \name changes when the value of $\lambda$ is set to $0.0$, while keeping all other settings unchanged.
For the ablation study, we conducted the experiments based on the hyperparameter configuration that demonstrated the highest MRR for \name in RQ1-1 to simplify the analysis: $(\lambda, \alpha, \tau)$ is set to $(0.1, 0, \mathtt{dense})$ for \name with IRFL, $(0.1, 0, \mathtt{max})$ for \name with SBFL.


\begin{table}
  \caption{Performance evaluation of three variants of \name with key features ablated: without Rank-based Voting Power, without Depth-based Decay, and without both features}
  \label{tab:ablation}
  \scalebox{0.90}{
  \begin{tabular}{l|rrr}
    \toprule
 &  MRR & Acc@1 & Acc@5 \\
    \midrule
      \textbf{Fonte with IRFL ($\alpha=0, \tau=\mathtt{dense}, \lambda=0.1$)} & \textbf{0.453} & \textbf{30.6\%} & \textbf{62.6\%} \\
      - w/o Ranking-based Voting Power & 0.404 & 24.8\% & 57.3\%\\
      - w/o Depth-based Decay& 0.422 & 27.7\% & 60.7\%\\
      - w/o Both & 0.378 & 24.3\% & 52.9\%\\\midrule
      \textbf{Fonte with SBFL ($\alpha=0, \tau=\mathtt{max}, \lambda=0.1$)} & \textbf{0.481} & \textbf{32.0\%} & \textbf{68.9\%} \\
      - w/o Ranking-based Voting Power & 0.446 & 29.1\% & 61.7\%\\
      - w/o Depth-based Decay & 0.447 & 29.1\% & 64.6\%\\
      - w/o Both & 0.414 & 27.7\% & 55.3\%\\\bottomrule
    \end{tabular}}
\end{table}

\noindent \textbf{Results} \Cref{tab:ablation} presents the performance evaluation of \name with key features ablated. The absence of Ranking-based Voting Power leads to a 10.8\% decrease in performance for \name-IRFL, and a 7.3\% decrease for \name-SBFL, in terms of MRR. Similarly, removing Depth-based Decay causes a 6.8\% performance drop with \name-IRFL and a 7.3\% drop with \name-SBFL. Eliminating both features results in a reduction of MRR by 16.6\% for \name-IRFL and 13.9\% for \name-SBFL. Specifically, for \name combined with IRFL, the lack of Ranking-based Voting Power is particularly detrimental to performance, while the removal of Depth-based Decay, although less impactful, still results in a significant performance decline. This pattern is consistent when \name is combined with SBFL, especially in terms of Acc@5. Collectively, these findings highlight the crucial role that both Ranking-based Voting Power and Depth-based Decay play in the overall performance of \name, with Rank-based Voting Power playing a more critical role in enhancing performance.

\noindent\textbf{Answer to RQ1-3}: The contribution of Rank-based Voting Power and Depth-based Decay to \name's efficacy is considerable: disabling both features results in up to a 16.6\% decrease in MRR.

\subsection{\textbf{RQ2. Comparison with Other Techniques}}

\noindent \textbf{Evaluation Protocol} We compare the BIC ranking performance of \name against various commit scoring baselines. To guarantee a fair comparison focused solely on the scoring mechanisms of each method and \name, we apply the baseline techniques within the same reduced BIC search space, $C_{BIC}$. The baselines include a general FL score aggregation method and two state-of-the-art IR-based techniques:

\begin{itemize}[leftmargin=1em]
\item Max Aggregation of FL Scores: In Orca~\cite{Bhagwan2018}, the file-level FL scores are converted into the commit level using max-aggregation, that is, the highest FL score among all files changed by the commit is assigned as the commit's score. This max-aggregation method is widely utilised in FL techniques to bridge differing granularities between initial FL scores and the targeted FL granularity~\cite{Sohn2017, Lou2020}, such as from statements to methods or files to components. To illustrate and evaluate this concept, we adopt a scoring model where the score of a commit is determined by the maximum FL score among the code elements it alters:
\begin{equation}
  commitScore(c) = \max_{e \in E_{F}^c}susp(e)
\label{eq:score_max}
\end{equation}

\item Bug2Commit~\cite{Murali2021}: Bug2Commit is a state-of-the-art IR-based
BIC identification method for large-scale systems, leveraging various
features of commits and bug reports. In our implementation of Bug2Commit, we opt for the Vector Space Model (VSM) because using a word-embedding model would necessitate an extra dataset comprising bug reports and commits for training. Following the approach detailed in the original study, we employ BM25~\cite{robertson1995okapi} for vectorisation. For the tokenisation process, the Ronin tokeniser is selected, recognised as the most sophisticated option available in \texttt{Spiral}~\cite{spiral2018}\footnote{https://github.com/casics/spiral}. 
We consider two features from each commit: the text of the commit message and the names of files that were altered. To represent bug reports, we evaluate two distinct configurations:
\begin{itemize}
\item Bug2Commit$_{report}$: Utilizes (1) the title and (2) the content of the bug report, both crafted by humans.
\item Bug2Commit$_{report+symptom}$: Builds on the previous configuration by additionally incorporating (3) the observed failure symptoms, such as exception messages and stack traces from failed test cases.
\end{itemize}

\item FBL-BERT~\cite{Ciborowska2022}: FBL-BERT is a recently proposed changeset
localisation technique based on a pre-trained BERT model called
BERTOverflow~\cite{tabassum2020code}. Given a bug report, it retrieves the
relevant changesets using their scores obtained by the BERT-based model. We
fine-tune the model using the training dataset from the \texttt{JDT} project,
which is the largest training dataset provided by their repository\footnote{We
confirm that the model fine-tuned with \texttt{JDT} performs better than that
fine-tuned with \texttt{ZXing}, which has the smallest training dataset.}: this
is because no such training data is available for our target projects. We use
the ARC changeset encoding strategy, which categorises the lines in the changeset into Added, Removed, and Context groups. This method has been demonstrated to perform the best for
changeset-level retrieval in the original study~\cite{Ciborowska2022}. As
Defects4J contains the link to the original bug report for every bug, we use
the contents of the original bug report as an input query.
\end{itemize}

\begin{table}
  \caption{Comparision of \name against commit scoring baselines. The methods marked with \dag utilise bug reports as input.}
  \label{tab:comparision}
  \centering
  \scalebox{1.00}{
  \begin{tabular}{l|rrr}
    \toprule
 &  MRR & Acc@1 & Acc@5 \\
    \midrule
      \textbf{Fonte with SBFL} & \textbf{0.481} & \textbf{32.0\%} & \textbf{68.9\%}\\
      \textbf{Fonte with IRFL} \dag & \textbf{0.453} & \textbf{30.6\%} & \textbf{62.6\%}\\\midrule
      Bug2Commit$_{report+symptom}$ & 0.330 & 17.0\% & 57.3\%\\
      Bug2Commit$_{report}$ \dag & 0.318 & 17.5\% & 54.4\% \\
      Max Aggregation of SBFL Scores & 0.288 & 12.1\% & 48.5\% \\
      Max Aggregation of IRFL Scores \dag & 0.274 & 11.7\% & 46.6\% \\
      FBL-BERT \dag & 0.246 & 14.1\% & 38.3\% \\\bottomrule
    \end{tabular}}
\end{table}

\noindent \textbf{Results} \Cref{tab:comparision} presents the ranking performance of the commit scoring baselines, sorted in descending order of MRR, along with the performance of \name. Among the baselines, Bug2Commit performs the best when utilising both the human-written bug report and failure symptoms as input to find the relevant commit. However, the results show that \name outperforms all scoring baselines across every evaluation metric. Specifically, \name with SBFL and IRFL demonstrates at least 45.8\% and 37.3\% higher MRR, respectively, compared to the baselines. This superior performance of \name holds true for any hyperparameter configuration, as shown in \Cref{tab:fonte_results}.

Furthermore, although three baselines (marked with \dag) use the same input data, i.e., the bug report, \name with IRFL achieves significantly superior performance. This highlights the effectiveness of \name's approach in converting code-level FL scores to commit-level scores. This is particularly evident when comparing \name with the max aggregation scheme, as they differ only in how the initial IRFL scores (in this case, Blues) are converted to the commit level.


\noindent\textbf{Answer to RQ2} The scoring approach of \name outperforms both the max-aggregation method for FL scores and the state-of-the-art IR-based techniques. Notably, when \name is paired with IRFL, which relies solely on the bug report for input, it achieves superior results compared to other methods that utilise the same input data.


\subsection{\textbf{RQ3. Standard Bisection vs. Weighted Bisection}}

\noindent \textbf{Evaluation Protocol} We simulate the standard and weighted bisection algorithms on all target bugs, assuming that the bug-revealing tests can perfectly reveal the existence of bugs. As commit scores, we use the best-performing configuration found in RQ1: \name with SBFL ($\alpha= 0$, $\tau=\mathtt{max}$, $\lambda=0.1$). We report how many search iterations until finding the BIC can be saved by using the weighted bisection algorithm compared to the standard bisection.

\begin{figure}[t]
  \centerline{\includegraphics[width=\linewidth]{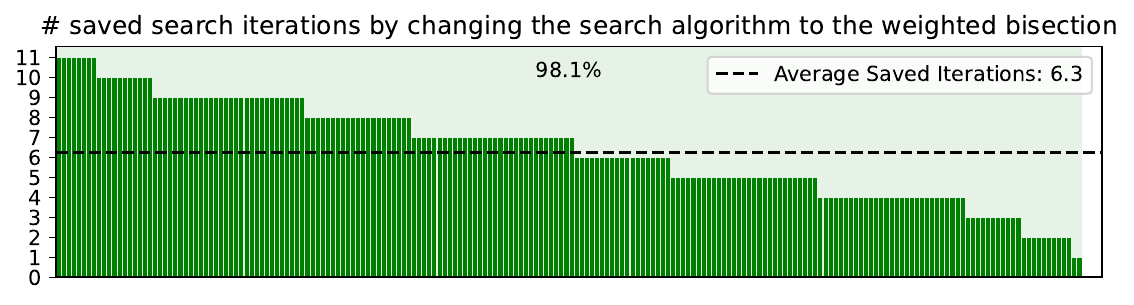}}
  \caption{The number of saved search iterations required until finding the BIC using the weighted bisection compared to the standard bisection on the \textbf{entire} commit history, $C$}
  \label{fig:RQ2-all}
\end{figure}

\begin{figure}[t]
  \centerline{\includegraphics[width=\linewidth]{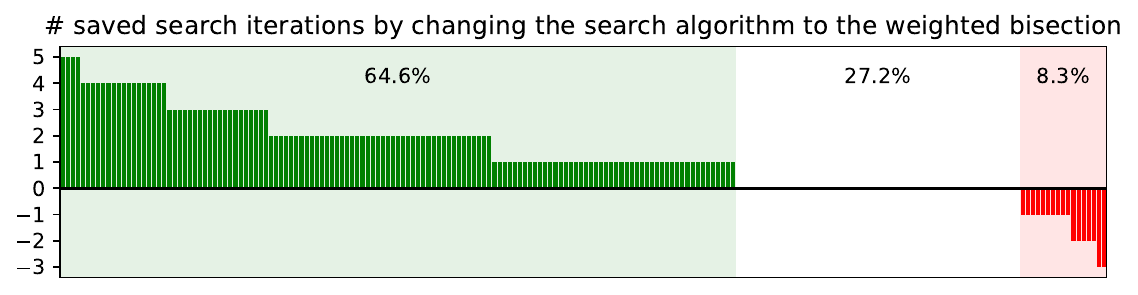}}
  \caption{The number of saved search iterations required until finding the BIC using the weighted bisection compared to the standard bisection on the \textbf{reduced} commit history, $C_{BIC}$}
  \label{fig:RQ2-reduced}
\end{figure}

\noindent \textbf{Results} \Cref{fig:RQ2-all} presents a sorted bar chart that shows the number of saved search iterations for all subjects until finding the BIC using the weighted bisection algorithm with \name-generated scores, compared to the standard bisection search on the entire commit history. The weighted bisection can reduce the search cost for approximately 98\% (202 out of 206) of the cases, saving up to 11 search iterations. On average, the number of iterations is reduced by 6.26, resulting in only 40\% of the iterations required by the standard bisection approach. Notably, there is no case where the weighted bisection degrades the performance compared to the standard bisection.

For a more conservative comparison, the weighted bisection algorithm using \name-generated scores is compared against the standard bisection search when both are applied to the reduced search space, $C_{BIC}$. \Cref{fig:RQ2-reduced} shows that the weighted bisection can reduce the number of required search iterations for 133 out of 206 subjects (64.6\%), while the number of iterations is increased in only 17 out of 206 subjects (8.3\%). In the remaining 27.2\% of cases, the number of iterations is the same as the standard bisection. The results demonstrate that the commit score information can guide the search process more efficiently. To ensure that the median of the number of saved iterations is positive, indicating a performance improvement, a one-sided Wilcoxon signed-rank test~\cite{Wilcoxon1992} is performed. The null hypothesis is that the median is negative, implying performance degradation. The obtained p-value of $1.69*10^{-19}$ allows the rejection of the null hypothesis in favor of the alternative that \emph{the median of the number of saved iterations is greater than zero, supporting the performance improvement}.

\begin{figure}[t]
  \centerline{\includegraphics[width=\linewidth]{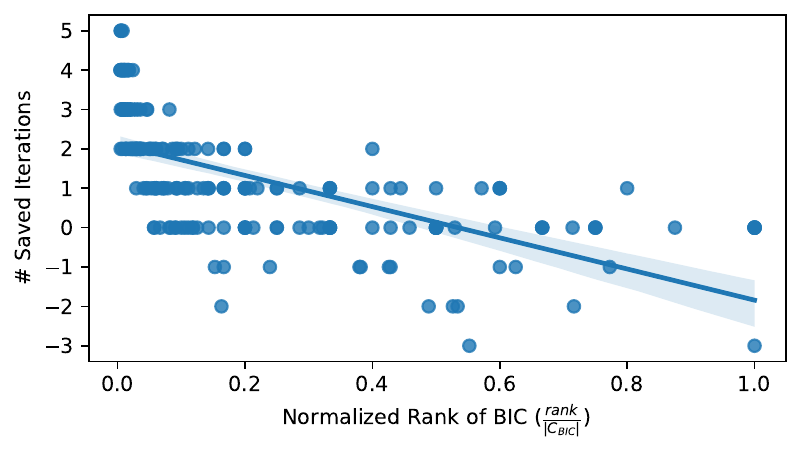}}
  \caption{Regression plot illustrating the relationship between the effectiveness of \name, measured by the normalized rank of the BIC, and the effectiveness of the weighted bisection technique, measured by the number of iterations saved compared to the standard bisection approach}
  \label{fig:RQ3-bic-rank-and-wb}
\end{figure}

We conducted an additional analysis to understand why the weighted bisection technique degrades the search efficiency for 17 subjects (8.3\%). In these cases, we found that the BIC was not ranked well by \name, either not being among the top 10 or even the top 50\% candidates. \Cref{fig:RQ3-bic-rank-and-wb} illustrates that the number of saved search iterations by using the weighted bisection exhibits a negative correlation with the normalised BIC rank, with a Pearson correlation coefficient of -0.64. This finding collectively suggests that more accurate commit scores can benefit the search process of the weighted bisection.

\noindent\textbf{Answer to RQ3} The combination of weighted bisection and \name-generated commit scores can reduce the BIC search cost for 98\% of the studied bugs, compared to the standard bisection applied to the entire commit history. On average, it saves 6.3 iterations. When the bisection is performed solely on the reduced set of candidate commits, the weighted bisection technique saves the number of search iterations in 65\% of cases, while increasing it in only 8\% of cases where the commit scores are of low quality.

\section{Application to Industry Software}
\label{sec:industry}

\product is a large-scale
commercial software that consists of more than 10M lines of C++ and C. In the
CI system of \product, multiple commits that have individually passed the
pre-submit testing are merged into the delivery branch and tested together using a more extensive test suite on a daily basis. Considering
the set of multiple commits as a single batch, this is a type of \emph{Batch
Testing}~\cite{Najafi2019}. While batch testing reduces the overall test
execution cost for \product, it also has some practical drawbacks: when a test
fails, it is not immediately clear which change in the batch is responsible for
the failure~\cite{Beheshtian2022}. The current CI system of \product identifies the BIC in the batch
using automatic bisection to aid the bug assignments~\cite{Bach2022}. However, each individual
inspection during the bisection can take up to several hours due to the
compilation, installation, and test execution cost, resulting in severe
bottlenecks in the overall debugging process. The bottleneck can be
particularly problematic if integration or system-level tests fail.

\begin{figure}[h]
  \centerline{\includegraphics[width=\linewidth]{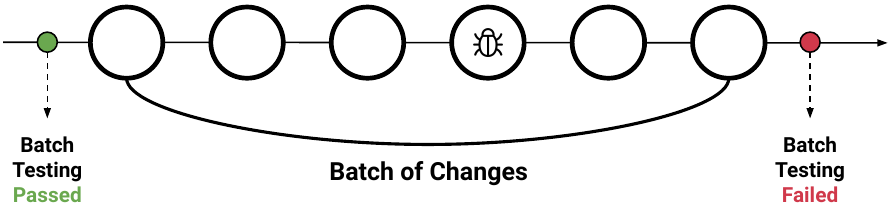}}
  \caption{Simplified batch testing scenario}
  \label{fig:batch_testing}
\end{figure}

This motivates us to see whether \name and its weighted bisection can reduce
the number of bisection iterations. To evaluate the effectiveness of
applying \name, we collect 23 batch testing failures that occurred from July to
August 2022 and their BICs identified by the bisection from the internal CI
logs of \product. Using the data, we first check if \name can find the BIC
inside the batch accurately (\Cref{fig:batch_testing}). As the test
coverage of \product is regularly and separately updated instead of being
measured at each of the batch testings, we use the latest line-level coverage
information to calculate the Ochiai scores. Note that we do not need to compute
the Ochiai scores for all lines, but only the lines covered by the failing
tests. When applying \name, depth-based voting decay is not used
($\lambda = 0$) because all candidate commits are submitted on the same day and
have not yet been merged into the main codebase. For the other
hyperparameters, we use $\alpha=1$ and $\tau=\mathtt{max}$ which performed the best with $\lambda=0$ in our experiment with Defects4J.

\begin{table}[t]
  \centering
  \caption{Evaluation of \name on the 23 batch testing failures of \product}
  \scalebox{1.00}{
    \begin{tabular}{l|r|rrrrr}
        \toprule
     &     \multirow{2}{*}{MRR} & \multicolumn{5}{c}{Accuracy}\\\cmidrule{3-7}
     &           &     @1 &     @2 &     @3 &     @5 &     @10 \\\midrule
     \multirow{2}{*}{\name}     &   \multirow{2}{*}{0.600}   & 10  & 14 & 15 & 20 & 23\\
             &    & (43\%) & (61\%) & (65\%) & (87\%) & (100\%) \\\midrule
    Random & 0.110 & 0 & 0 & 0 & 1 & 17\\\bottomrule
    \end{tabular}}
  \label{tab:industry}
\end{table}

\Cref{tab:industry} shows the BIC ranking performance of \name in terms of
MRR and Accuracy@n. While each batch contains 18.48 commits on average, \name
can locate the actual BIC within the top 1 and 5 for 43\% and 87\% of the
failures, respectively. Compared to the random baseline, it achieves 5.5-fold
increase in MRR.
Further, we also report that the weighted bisection can reduce the
bisection iterations for 18 out of 23 cases (78\%), while it increases the cost
in only three cases (13\%). Based on this result, we plan to incorporate weighted
bisection into the CI process of \product, which is expected to save 32\% of
required iterations. Considering that each iteration can take up
to several hours, we expect a significant reduction in the average BIC
identification cost for \product in the long run.

\section{Threats to Validity}
\label{sec:threats}

Threats to internal validity concern factors that can affect how confident we
are about the causal relationship between the treated factors and the effects.
\name relies on widely-adopted open-source tools to establish $\mathsf{Cover}$ and $\mathsf{Evolve}$
relations to ensure the chain of causality between the test failure and BIC identification.
We compare the validity of these different code history mining tools, that are used to build the $\mathsf{Evolve}$ relation, and use the most reliable one, \texttt{git log}, for the later experiments. We also make the performance results with the other tools publicly available in our artifact for future scrutiny, Additionally, as the baseline techniques rely on multiple sources of 
information, such as bug reports, we choose Defects4J as our benchmark as it provides 
well-established links between real-world bug reports and the buggy version, not to mention
human-written bug-revealing test cases that withheld scrutiny from the community.

Threats to external validity concern factors that may affect how well our findings
can be generalised to unseen cases. Our key findings are primarily based on experiments with
the open-source Java programs in Defects4J. Since they are not representative of
the entire population of real-world programs, only further evaluations can strengthen
our claim of generalisation. We tried to support our claim by evaluating \name with
industry-scale software written in C and C++. We do note that \name does not generalise
to bugs that are caused by non-executable files, such as configuration changes, as its
base assumption is that the test failure is caused by a bug in the source code.
We leave extension of \name to bugs caused by non-executable changes as our primary
future work.

Threats to construct validity concern how well the used metrics measure the properties
we aim to evaluate. We adopt two ranking-evaluation metrics, MRR and Accuracy@n, to
evaluate \name: both have been widely used in the IR and SE literature. Since they are based on absolute ranks, we do note that the results can be overrated
when the number of ranking candidates is small. To mitigate the threat, we also
present the expected and worst values for the measures as baselines.

\section{Related Work}
\label{sec:related_work}

Locus~\cite{Wen2016} is the first work that proposed to localise the bug at the
software change level. It takes a bug report as an input query and locates the
relevant change hunk based on the token similarities. IR-based techniques, such
as Locus, and \name can complement each other depending on circumstances. When
the failure cannot be reproduced from the bug report, IR-based techniques can
be used instead of \name. However, if the coverage of the failing and passing
tests are available, we can apply \name with SBFL to more precisely rank the
commits without relying on IR.
ChangeLocator~\cite{Wu2017} aims to find a BIC for crashes using the call stack
information. It is a learning-based approach that requires data from fixed
crashes. Unlike ChangeLocator, \name is not limited to crashes and can be
applied to general failures. Orca~\cite{Bhagwan2018} takes symptoms of bugs,
such as an exception message or customer complaints, as an input query and
outputs a ranked list of commits ordered by their relevance to the query. It
uses the TF-IQF~\cite{Yang2008} to compute the relevance scores of files, and
aggregate them to a commit level. Subsequently, it uses machine learning to
predict the risk of candidate commits for breaking ties.
Bug2Commit~\cite{Murali2021} uses multiple features extracted from bug reports
and commits, and aggregates all features by taking the average of their vector
representations. Although Bug2Commit uses an unsupervised learning approach, it
needs the historical data of project-specific bug reports and commits to train
the word embedding model. FBL-BERT~\cite{Ciborowska2022} retrieves the relevant
changeset for the input bug report using a fine-tuned BERT model that can
capture the semantics in the text. It proposes fine-grained changeset encoding methods and accelerates the retrieval by offline indexing~\cite{johnson2019billion}. The major difference between \name and the techniques
mentioned above is that \name does not require any training. Further, \name can
be combined with any code-level FL technique, without being coupled to
specific sources of information, as long as the coverage of failing executions
is available.

The weighted bisection algorithm we propose is similar to FACF (Flaky Aware Culprit Finding)~\cite{Henderson2023}, which formulates the \emph{flake-aware} bisection problem as a Bayesian inference, in that both guide the bisection process based on the probability of commits being a source of test failure. The difference between the two algorithms is that ours uses commit scores from \name to establish the initial probability distribution, while  FACF updates the probability based on the test results during the search taking into account the potential for flakiness. The original work notes that FACF can take into account any prior information about the bug inducing change in the form of an initial probability distribution. Hence, we believe that the commit scores generated by \name can be used as an effective prior distribution for the FACF framework.

There exist studies that are highly relevant to \name despite not being specifically about the BIC identification domain. FaultLocator~\cite{zhang2011localizing} is similar to \name as both use
code-level FL scores to identify suspicious changes. FaultLocator combines spectrum information with the change impact
analysis to precisely identify the failure-inducing \emph{atomic} edits out
of all edits between two versions, whereas \name aims to pinpoint BICs in the
commit history. WhoseFault~\cite{Servant2012} is a method that utilises code-level FL scores and commit history to determine the developer responsible for a bug. While it provides insights into the assignment of bugs, it does not specifically target BIC identification. As a result, it cannot be directly compared with \name in our evaluation, nor can it be integrated with our bisection algorithm. Our belief is that accurately identifying the BIC can also be used to find the developer responsible for fixing the bug, based on the authorship of the changes, in addition to helping developers understand the context in which the failure occurred.

\section{Conclusion}
\label{sec:conclusion}

This paper proposes \name, a BIC identification technique that is available upon the observation of a failure.
It prunes the BIC search space using failure coverage and the syntactic analysis of commits, and assigns scores
to the remaining commits using the FL scores as well as change histories of code elements. Our experiments with
206 bugs in Defects4J show that \name can effectively identify BICs with an MRR of 0.481, which significantly
outperforms the baselines including state-of-the-art BIC identification techniques. The findings indicate that \name is an effective algorithm to translate code-level suspiciousness scores into commit-level scores. Along with \name, we also
propose the weighted bisection to accelerate the BIC search utilising the commit score information and show
that it can save the search cost in 98\% of the studied cases compared to the standard bisection. Finally,
the application of \name to a large-scale industry software \product shows that \name can successfully
reduce the cost of BIC identification in a batch-testing CI scenario. 


\bibliographystyle{IEEEtran}
\bibliography{references}


 




\vfill

\end{document}